\documentclass[showpacs,aps,prb,amsmath,amssymb]{revtex4}
\usepackage{graphicx}
\usepackage{dcolumn}
\usepackage{bm}
\begin{document}

\renewcommand{\thefigure}{\arabic{figure}}
\title{Analytic theory of ground-state properties of a three-dimensional electron gas 
at varying spin polarization}
\author{B. Davoudi,$^{1,2}$ R. Asgari,$^{1,2}$ M. Polini,$^{1}$ and M. P. Tosi$^{1}$}
\affiliation{$^1$NEST-INFM and Classe di Scienze, Scuola Normale Superiore, I-56126 Pisa, Italy\\
$^2$Institute for Studies in Theoretical Physics and Mathematics, Tehran
19395-5531, Iran\\}
\begin{abstract}
We present an analytic theory of the spin-resolved pair distribution functions $g_{\sigma\sigma'}(r)$ and the ground-state energy of an electron gas with an arbitrary degree of spin polarization. We first use the Hohenberg-Kohn variational principle and the von Weizs\"{a}cker-Herring ideal kinetic energy functional to derive a zero-energy scattering Schr\"{o}dinger equation for $\sqrt{g_{\sigma\sigma'}(r)}$. The solution of this equation is implemented within a Fermi-hypernetted-chain approximation which embodies the Hartree-Fock limit and is shown to satisfy an important set of sum rules. We present numerical results for the ground-state energy at selected values of the spin polarization and for $g_{\sigma\sigma'}(r)$ in both a paramagnetic and a fully spin-polarized electron gas, in comparison with the available data from Quantum Monte Carlo studies over a wide range of electron density. 
\end{abstract}
\maketitle

\section{Introduction}
\label{sec1}
The homogeneous electron gas (EG) of fermions interacting by the $e^2/r$ law and moving in a uniform neutralizing background of positive charge has been for many decades the basic reference system for calculations of electronic structure in condensed matter physics.\cite{ceperley,march} The extent to which exchange and correlations compete with each other on varying the electron density has been a central issue going back to the work of Bloch on spontaneous spin polarization due to exchange and to that of Wigner and Seitz on cohesion in simple metals. The many-body effective interactions acquire a magnetic component arising from local deviations of electron density around any given electron, with the exchange term being determined by the electrons carrying the same spin and the correlation term being to a large extent determined by the electrons of opposite spin. The role of spin fluctuations and their correlations thus are a subject of continuing interest, on which theoretical progress starting from the basic Coulomb Hamiltonian has been painfully slow. 

There is a broad consensus on the increasing importance of spin polarization at strong coupling and of correlations in lowered dimensionality. This fact has been repeatedly emphasized by Quantum Monte Carlo (QMC) numerical studies, starting from the early work of Ceperley and Alder.\cite{ca} In particular, in the three-dimensional $(3D)$ electron system these studies have revealed a continuous transition from a paramagnetic to a fully spin-polarized ground state occurring with increasing coupling strength, before a first-order quantum phase transition to a ferromagnetic Wigner crystal occurs.\cite{ortiz,zong} While the $3D$ system appears to be most suitable at present for improvements and critical tests of analytic theories, it may be mentioned that studies of spin motions in structures of reduced dimensionality are being greatly stimulated by expectations of developments in spin-based electronics and in quantum computing.\cite{dms}

The main purpose of the present work is the formulation of a practicable analytic theory for spin correlations in the $3D$ EG and the comparison of its numerical predictions with the available QMC data. The fluid that we consider consists of two spin components at fixed average densities $n_{\sigma}$  with $\sigma =\uparrow$ or $\downarrow$, and the spin correlations are described by the pair distribution functions $g_{\sigma\sigma'}(r)$. These are defined so that the quantity $4\pi r^2 n_{\sigma'}g_{\sigma\sigma'}(r)dr$  gives the average number of electrons with spin $\sigma'$ lying within a spherical shell of radius $r$ and thickness $dr$ centered on an electron with spin $\sigma$. The general strategy is to set up a Schr\"{o}dinger-like differential equation for $\sqrt{g_{\sigma\sigma'}(r)}$ with the help of the Hohenberg-Kohn variational principle\cite{d&g} and to solve it by a Fermi-hypernetted-chain type of approximation tailored to embody the Hartree-Fock limit as well as a set of sum rules for the two-component Fermi fluid. 

It may be mentioned at this point that there has quite recently been a renewed interest in the study of the spin-averaged pair distribution function $g(r)$ in paramagnetic EG models within a two-body scattering approach first proposed by Overhauser.\cite{gr} In brief, a Schr\"{o}dinger equation is set up to evaluate particle-pair wave functions from which $g(r)$ can be built through sums over angular momentum and spin. Various approximations have been examined for embodying the many-body effects into the effective scattering potentials.\cite{gr} In an alternative approach, which is fully developed in the present work, we have instead sought to use a density-functional framework for a self-consistent evaluation of $g(r)$ and of the effective scattering potential.\cite{dbose}
	
The contents of the paper are briefly as follows. Section II presents a formally exact zero-energy scattering equation for the partial distribution functions $g_{\sigma\sigma'}(r)$  and introduces the approximations that we propose for the evaluation of the three main contributions to the scattering potential. Section III shows that these approximations satisfy the plasmon sum rule, the charge neutrality condition, and Kimball's cusp condition. Our numerical results for the $3D$ EG are presented in Section IV. Finally, Section V summarizes our main conclusions and gives suggestions for further work.

\section{Theory}
\label{sec2}
We consider an inhomogeneous $3D$ fluid of electrons consisting of the two spin species with densities $n_{\sigma}({\bf r})$ in the presence of external potentials $V^{\rm ext}_{\sigma}({\bf r)}$ and of a uniform neutralizing background. From the Hohenberg-Kohn theorem\cite{d&g} the ground-state energy functional of the fluid can be written as
\begin{equation}\label{e1}
E_{\rm gs}[\left\{n_{\sigma}({\bf r})\right\}]=T_s[\left\{n_{\sigma}({\bf r})\right\}]+\sum_{\sigma}\int\,d^3{\bf r}\,\,V^{\rm ext}_{\sigma}({\bf r})\Delta n_{\sigma}({\bf r})+E_{\scriptscriptstyle \rm H}[\left\{n_{\sigma}({\bf r})\right\}]+E_{\rm xc}[\left\{n_{\sigma}({\bf r})\right\}]
\end{equation}
where $\Delta n_{\sigma}({\bf r})=n_{\sigma}({\bf r})-n_{\sigma}$ are the deviations of the spin densities from their average values, $T_s$ and $E_{\rm xc}$ are the ideal kinetic energy and exchange-correlation energy functionals, and $E_{\scriptscriptstyle \rm H}$ is the Hartree term given by
\begin{equation}\label{e2}
E_{\rm H}[\left\{n_{\sigma}({\bf r})\right\}]=\frac{1}{2}\sum_{\sigma,\sigma'}\int d^3{\bf r}\int d^3{\bf r}'\,v(|{\bf r}-{\bf r}'|)\,\Delta n_{\sigma}({\bf r})\Delta n_{\sigma'}({\bf r}')
\end{equation}
with $v(|{\bf r}-{\bf r}'|)=e^2/|{\bf r}-{\bf r}'|$. The presence of a neutralizing background has been taken into account in these equations. Following Herring\cite{herring} we decompose the kinetic energy functional into the sum of two terms,
 
\begin{equation}\label{e3}
T_s[\left\{n_{\sigma}({\bf r})\right\}]=\frac{\hbar^2}{8 m}\,\sum_{\sigma} \int d^3{\bf r}\,
\frac{|\nabla n_{\sigma}({\bf r})|^2}{n_{\sigma}({\bf r})} +T_{\theta}[\left\{n_{\sigma}({\bf r})\right\}]
\end{equation}
where the first term is the von Weizs\"acker "surface" kinetic energy\cite{w} and 
$T_{\theta}$ is itself defined by Eq. (3).
 
This general formalism is adapted to the derivation of differential equations for the spin-resolved pair distribution functions $g_{\sigma\sigma'}(r)$ in the {\it homogeneous} EG by viewing the quantity $n_{\sigma}\,[g_{\sigma\sigma'}(r)-1]$ as the distortion that an electron in the EG, located at position 
$r=0$ with spin $\sigma'$, induces in the density profiles of the electrons with spin $\sigma$.\cite{percus} 
The appropriate ground-state energy functional for this problem is obtained from Eqs. (1)-(3) 
by the formal replacements
\begin{equation}\label{e4}
\left\{ \begin{array}{ll}
											m & \rightarrow \mu \\
                      V^{\rm ext}_{\sigma}({\bf r}) & \rightarrow  v(r) \\
											n_{\sigma}({\bf r}) & \rightarrow  
											n_{\sigma}\,g_{\sigma \sigma'}(r)
				\end{array}
				\right.							
\end{equation}
where $\mu=m/2$ is the reduced mass of an electron pair. The Euler-Lagrange equations for the spin-resolved pair functions can now be obtained from the variational principle of Hohenberg and Kohn\cite{d&g} using the 
von Weizs\"acker-Herring ideal kinetic energy functional as shown in Eq. (3). With the zero of energy taken at the chemical potential, the formally exact differential equation for $g_{\sigma\sigma'}(r)$ reads
\begin{equation}\label{e5}
\left[-\frac{\hbar^2}{m}\nabla^2_{\bf r}+v(r)+v^{\sigma\sigma'}_{\rm \scriptscriptstyle P}(r)+V^{\sigma\sigma'}_{\scriptscriptstyle \rm ex}(r)\right]\sqrt{g_{\sigma\sigma'}(r)}=0\,.
\end{equation}
Here, the "Pauli potential" $v^{\sigma\sigma'}_{\rm \scriptscriptstyle P}(r)$ is defined by
\begin{equation}\label{e7}
v^{\sigma\sigma'}_{\rm \scriptscriptstyle P}(r)=\left.\frac{\delta T_{\theta}[\left\{n_{\sigma}({\bf r})\right\}]}{\delta n_{\sigma }({\bf r})}\right|_{n_{\sigma}({\bf r})=n_{\sigma}\,g_{\sigma \sigma'}(r)}
\end{equation}
and $V^{\sigma \sigma'}_{\scriptscriptstyle \rm ex}(r)$ 
is the "excess" Kohn-Sham potential, which is given by
\begin{equation}\label{e6}
V^{\sigma \sigma'}_{\scriptscriptstyle \rm ex}(r)= v_{\rm H}(r)+v^{\sigma\sigma'}_{\rm xc}(r)=n\int d^3{\bf r}' v(|{\bf r}-{\bf r}'|)[g(r')-1]+\left.\frac{\delta E_{\rm xc}[n_{\sigma\sigma'}({\bf r})]}{\delta n_{\sigma\sigma'}({\bf r})}\right|_{n_{\sigma}({\bf r})=n_{\sigma}\,g_{\sigma \sigma'}(r)}
\end{equation}
with $n=n_{\uparrow}+n_{\downarrow}$ and $g(r)=\sum_{\sigma, \sigma'}(n_{\sigma}n_{\sigma'}/n^2)g_{\sigma\sigma'}(r)$. 

Equations (5)-(7) show that a zero-energy scattering theory approach to pair correlations in a quantum electron fluid has a sound theoretical justification within the framework of the Hohenberg-Kohn variational principle. Of course, the functional dependence of $T_{\theta}$ and of $E_{\rm xc}$ on the inhomogeneous electron densities is not known and we shall need to resort to appropriate calculational schemes and approximations for the potentials introduced in Eqs. (6) and (7).

A possible way to handle the kinetic energy term is to pass to a Kohn-Sham scheme, by expanding the pair distribution functions into Kohn-Sham two-particle scattering orbitals. This was done in a series of previous studies\cite{gr,dbose} as already mentioned in Section I, but the model scattering potentials used there fail to satisfy the plasmon sum rule and therefore do not account correctly for the behavior of $g_{\sigma\sigma'}(r)$ at large $r$. In the following we shall instead take advantage of the formal similarity between Eq. (5) and the Euler-Lagrange equation which is adopted in the so-called Fermi hypernetted-chain approximation (FHNC). This is derived by a Jastrow-Feenberg variational {\it Ansatz} on the many-body wave function\cite{lantto,zab,report} and was used to treat two-component Fermi fluids in the case of electron-hole liquids and of liquid metallic hydrogen.\cite{tc} Such an approach will allow us to build into the theory some important sum rules and limiting behaviors.
\subsection{The Pauli potential}

In the FHNC context the Pauli potential can be chosen so as to ensure that the Hartree-Fock limit is correctly embodied into the theory. An argument can be given to show that this term in Eq. (5) becomes dominant at weak coupling by considering how the various functionals scale under scaling of all lengths by a factor $\lambda$. It was shown by Herring\cite{herring} that $T_{\theta}$ scales by a factor $\lambda^{-2}$, whereas Levy and Perdew\cite{levy} have shown that both the Hartree and the exchange energy functionals scale by a factor $\lambda^{-1}$. Only the inequality   $E_c[\{n_{\sigma}({\bf r})\}]>\lambda^{-1}E_c[\{n_{\sigma}^{(\lambda)}({\bf r})\}]$ could be proved\cite{levy} 
for the correlation energy functional under scaling, but it seems reasonable to expect that on taking $\lambda=r_s$ the correlation functional would also become relatively negligible in the limit $r_s \rightarrow 0$. Here $r_s$ is the usual coupling strength parameter for the EG, which in $3D$ is related to the average electron density $n$ by $r_s a_B=(4 \pi n/3)^{-1/3}$ with $a_B$ the Bohr radius. 	

From the above argument it follows that Eq. (5) implies
\begin{equation}\label{e8}
v^{\sigma\sigma'}_{\rm \scriptscriptstyle P}(r)
=\frac{\hbar^2}{m}~\frac{\nabla_{\bf r}^2 \sqrt{g^{\rm \scriptscriptstyle HF}_{\sigma\sigma'}(r)}}{\sqrt{g^{\rm \scriptscriptstyle HF}_{\sigma\sigma'}(r)}}
\end{equation} 
in the weak coupling limit $r_s \rightarrow 0$. In Eq. (8) $g^{\rm \scriptscriptstyle HF}_{\sigma\sigma'}(r)$ are the spin-resolved pair functions in the Hartree-Fock approximation (HF), namely
\begin{equation}
\left\{
\begin{array}{l}
g^{\rm \scriptscriptstyle HF}_{\sigma\sigma}(r)=1-9~[{\rm j}_1(k_{F\sigma} r)/(k_{F\sigma} r)]^2 \\
g^{\rm \scriptscriptstyle HF}_{\sigma {\bar \sigma}}(r)=1\\
\end{array}
\right.
\end{equation}
where ${\bar \sigma}=-\sigma$, ${\rm j}_1(x)=(\sin{x}-x\cos{x})/x^2$  is a spherical Bessel function, and $k_{F\sigma}=k_F(1+{\rm sgn}(\sigma)\zeta)^{1/3}$  with $k_F=(3 \pi^2 n)^{1/3}$ the usual Fermi wave number and $\zeta=|n_\uparrow-n_\downarrow|/n$ the degree of spin polarization. 	

Although the expression for the Pauli potential in Eq. (8) is correct only for a weakly coupled Fermi fluid, we shall assume in the following that it can yield useful results in our self-consistent calculations of the pair distribution functions with increasing coupling strength. This assumption will have to be tested through quantitative comparisons of our numerical results with the available QMC data. As a broad qualitative argument in support of this assumption we may remark that the role of the statistics is expected to weaken with increasing coupling strength at least in the spin-averaged pair function\cite{chiofalo,dharma} and that only the von Weizs\"acker term contributes to the kinetic energy functional in the case of a charged Bose fluid.\cite{dbose}

\subsection{The excess Kohn-Sham potential}

The FHNC expresses the potential $V^{\sigma\sigma'}_{\scriptscriptstyle \rm ex}(r)$ in Eq. (5), which is the sum of the Hartree and of the exchange-correlation potential, as the sum of two effective pair interactions:\cite{lantto,zab,report,tc}
\begin{equation}
V^{\sigma\sigma'}_{\scriptscriptstyle \rm ex}(r)=W^{\sigma\sigma'}_{\rm B}(r)+\delta_{\sigma\sigma'}W^{\sigma\sigma}_{\rm e}(r)\,.
\end{equation}
The first term on the left-hand side of Eq. (10) descends from the two-body correlation functions $u_{\sigma\sigma'}(r)$ in the Jastrow-Feenberg wave function and is therefore formally the same as for a binary boson mixture. The second term is instead due to the antisymmetry of the fermion many-body wave function.

As shown by Chakraborty\cite{tc} in treating a binary fermion mixture, the HNC closure yields for $u_{\sigma\sigma'}(r)$ the expression
\begin{equation}
u_{\sigma\sigma'}(r)=\ln{g_{\sigma\sigma'}(r)}-[g_{\sigma\sigma'}(r)-1]+c_{\sigma\sigma'}(r)
\end{equation}
where $c_{\sigma\sigma'}(r)$ are the direct correlation functions, which are related to $g_{\sigma\sigma'}(r)$ 
by the Ornstein-Zernike relations. We introduce at this point the partial structure factors $S_{\sigma\sigma'}(q)$ 
of the binary mixture, which in essence are the Fourier transforms of $g_{\sigma\sigma'}(r)$:
\begin{equation}\label{sq}
S_{\sigma\sigma'}(q)=\delta_{\sigma\sigma'}+\sqrt{n_{\sigma}n_{\sigma'}}\,\int d^{3} {\bf r}\,[g_{\sigma\sigma'}(r)-1]\exp{(-i {\bf q} \cdot {\bf r})}\,.
\end{equation}
We also introduce the Fourier transform of $W^{\sigma\sigma'}_{\rm B}(r)$  ($W^{\sigma\sigma'}_{\rm B}(q)$ , say). Minimization of the ground state energy against arbitrary variations of $g_{\sigma\sigma'}(r)$ yields with the help of Eq. (11) the expression
\begin{equation}\label{e8_bis}
W^{\sigma\sigma'}_{\rm B}(q)=-\frac{\varepsilon_{q}}{\sqrt{n_{\sigma}n_{\sigma'}}}
[S_{\sigma \sigma'}(q)-\delta_{\sigma\sigma'}]-V_{\sigma \sigma'}(q)
\end{equation}
where $\varepsilon_{q}=\hbar^2 q^2/(2m)$ are the single-particle kinetic energies and the functions $V_{\sigma \sigma'}(q)$ 
are given by
\begin{equation}
\left\{
\begin{array}{l}
V_{\sigma \sigma}(q)={\displaystyle \frac{\varepsilon_{q}}{2 n_{\sigma}}\left\{-1+[S^2_{{\bar \sigma}{\bar \sigma}}(q)+S^2_{\sigma {\bar \sigma}}(q)]/\Delta^2(q)\right\}}\\
V_{\sigma {\bar \sigma}}(q)={\displaystyle -\frac{\varepsilon_{q}}{2 \sqrt{n_{\sigma}n_{\bar \sigma}}}
S_{\sigma {\bar \sigma}}(q)\,\left[S_{\sigma\sigma}(q)+S_{{\bar \sigma} {\bar \sigma}}(q)\right]/\Delta^2(q)}
\end{array}
\right.
\end{equation}
with
\begin{equation}
\Delta(q)=S_{\sigma \sigma}(q)S_{{\bar \sigma} {\bar \sigma}}(q)-S^2_{\sigma {\bar \sigma}}(q)\,.
\end{equation}
Equations (13)-(15) show how the effective boson-like interactions $W^{\sigma\sigma'}_{\rm B}(r)$ in Eq. (10) are related in Fourier transform to the spin-resolved pair distribution functions. 	

Turning to the second term on the left-hand side of Eq. (10), the effective pair potential $W^{\sigma\sigma}_{\rm e}(r)$ has a very complicated expression within the FHNC.\cite{lantto,zab,report} However, in dealing with a one-component electron fluid Kallio and Piilo\cite{kp} have proposed a simple and effective way to account for this consequence of the antisymmetry of the fermion wave function. Their argument is immediately generalized to our two-component Fermi fluid, and leads to the requirement that in Fourier transform this term should cancel the effective boson-like interaction $W^{\sigma\sigma}_{\rm B}(q)$ for parallel-spin electrons at low coupling. That is,
\begin{equation}\label{we}
W^{\sigma \sigma}_{\rm e}(q)=-\lim_{r_s \rightarrow 0}W^{\sigma \sigma}_{\rm B}(q)=
\frac{\varepsilon_q}{2 n_{\sigma}}[1+2~S^{\rm \scriptscriptstyle HF}_{\sigma\sigma}(q)]\left[\frac{S^{\rm \scriptscriptstyle HF}_{\sigma\sigma}(q)-1}{S^{\rm \scriptscriptstyle HF}_{\sigma\sigma}(q)}\right]^2~.
\end{equation}
Here, $S^{\rm \scriptscriptstyle HF}_{\sigma\sigma}(q)$ is the Hartree-Fock structure factor, which is given by
\begin{equation}
S^{\rm \scriptscriptstyle HF}_{\sigma\sigma}(q)=
\left\{
\begin{array}{l}
\frac{3}{4}(q/k_{F\sigma})-\frac{1}{16}(q/k_{F\sigma})^3\,\,\mbox{for $q\leq 2 k_{F\sigma}$}\\
1\,\,\,\,\hspace{3.3 cm} \mbox{for $q\geq 2 k_{F\sigma}$}.
\end{array}
\right.
\end{equation}

It is evident that the insertion of Eqs. (8)-(17) into Eq. (5) allows a self-consistent calculation of the spin-resolved pair distribution functions and of the effective electron-electron interactions. Before proceeding to the numerical solution of this problem, we examine how the approximate theory presented above fares in regard to some exact properties of pair correlations. 

\section{Sum Rules and Limiting behaviors}
In this Section we show that the pair functions obtained from the theory presented in Section II satisfy three exact properties. These are the charge neutrality condition, the plasmon sum rule, and Kimball's cusp condition. 

The asymptotic behavior of the effective potential $V_{\sigma \sigma'}(r)$ is first obtained from Eq. (5), which can be rewritten in the form
\begin{equation}\label{ph}
V_{\sigma\sigma'}(r)=g_{\sigma\sigma'}(r)[v(r)+W^{\sigma\sigma'}_{\rm e}(r)+v^{\sigma\sigma'}_{\rm \scriptscriptstyle P}(r)]+[g_{\sigma\sigma'}(r)-1]W^{\sigma\sigma'}_{\rm B}(r)+\frac{\hbar^2}{2 \mu}\left|\nabla \sqrt{g_{\sigma\sigma'}(r)}\right|^2~.
\end{equation}
A careful analysis of this equation shows that $V_{\sigma\sigma'}(r)\rightarrow v(r)+\delta_{\sigma\sigma'}W^{\sigma\sigma}_{\rm e}(r)$ for $r \rightarrow \infty$, 
and hence in Fourier transform we have from Eq. (16)
\begin{equation}
\left\{
\begin{array}{l}
V_{\sigma\sigma}(q)\rightarrow v_q+8 \pi^2 \hbar^2/(3 m k_{F\sigma})\\
V_{\sigma {\bar \sigma}}(q)\rightarrow v_q
\end{array}
\right.
\end{equation}
for $q \rightarrow 0$, with $v_q= 4 \pi e^2/q^2$ in the $3D$ EG. The corresponding asymptotic behavior of the structure factors is obtained from Eqs. (14), which can be inverted to yield
\begin{equation}
\left\{
\begin{array}{l}
S_{\sigma\sigma}(q)={\displaystyle \sqrt{A_{{\bar \sigma} {\bar \sigma}}\Delta^2-\frac{A^2_{\sigma{\bar \sigma}}\Delta^3}{2+[A_{\sigma\sigma}+A_{{\bar \sigma} {\bar \sigma}}]\Delta}}}\\
S_{\sigma{\bar \sigma}}(q)={\displaystyle \frac{A_{\sigma{\bar \sigma}}\Delta^{3/2}}{\sqrt{2+[A_{\sigma\sigma}+A_{{\bar \sigma} {\bar \sigma}}]\Delta}}}
\end{array}
\right.
\end{equation}
where $A_{\sigma\sigma'}(q)=\delta_{\sigma\sigma'}+2\,{\rm sgn}{(\sigma\sigma')}\sqrt{n_{\sigma}n_{\sigma'}}\,V_{\sigma\sigma'}(q)/\varepsilon_q$ and $\Delta(q)=(A_{\uparrow\uparrow}A_{\downarrow\downarrow}-A^2_{\uparrow\downarrow})^{-1/2}$. 
From Eqs. (19) and (20) we find
\begin{equation}
S_{\sigma \sigma'}(q)\rightarrow f_{\sigma\sigma'}(\zeta)\frac{q}{k_F}+g_{\sigma\sigma'}(\zeta)
\frac{\varepsilon_q}{2 \hbar \omega_{\rm pl}}
\end{equation}
for $q\rightarrow 0$, with $\omega_{\rm pl}=(4 \pi n e^2/m)^{1/2}$, $g_{\sigma\sigma}(\zeta)=[1+{\rm sgn}(\sigma)\zeta]$, $g_{\sigma {\bar \sigma}}(\zeta)=\sqrt{1-\zeta^2}$, and
\begin{equation}
\left\{
\begin{array}{l}
f_{\sigma\sigma}(\zeta)={\displaystyle \frac{3}{8}\sqrt{\frac{2[1-{\rm sgn}(\sigma)\zeta]/[1+{\rm sgn}(\sigma)\zeta]}{(1+\zeta)^{-1/3}+(1-\zeta)^{-1/3}}}}\\
f_{\sigma {\bar \sigma}}(\zeta)=-{\displaystyle \frac{3}{8}\sqrt{\frac{2}{(1+\zeta)^{-1/3}+(1-\zeta)^{-1/3}}}}\,.
\end{array}
\right.
\end{equation}
These limiting expressions were earlier known to hold at $\zeta=0$ 
within the Random Phase approximation (see e.g. Gori-Giorgi {\it et al.}\cite{gsb} and references given therein). 
The form of Eq. (21) immediately ensures that the charge neutrality condition is satisfied. This reads
\begin{equation} 
\sqrt{n_{\sigma} n_{\sigma'}}~\int d^3 {\bf r}~[g_{\sigma\sigma'}(r)-1]=-\delta_{\sigma\sigma'}
\end{equation}
or $S_{\sigma\sigma'}(q\rightarrow 0)=0$ from Eq. (12).

The form of Eq. (21) also ensures that the plasmon sum rule is satisfied. This reads
\begin{equation}
\lim_{q\rightarrow 0} S(q)=\varepsilon_q/(\hbar \omega_{\rm pl})
\end{equation}
where $S(q)=\sum_{\sigma,\sigma'}(\sqrt{n_{\sigma} n_{\sigma'}}/n)S_{\sigma\sigma'}(q)$.

The cusp condition as first demonstrated by Kimball\cite{kimball} relates the logarithmic derivative of $g_{\uparrow \downarrow}(r)$ in the origin to the Bohr radius and is satisfied if the scattering potential between pairs of electrons with antiparallel spins reduces to the bare Coulomb potential for $r \rightarrow 0$. 
This is ensured in our approach by the properties $v^{\sigma\sigma}_{\rm \scriptscriptstyle P}(r\rightarrow 0)=W^{\sigma\sigma'}_{\rm e}(r \rightarrow 0)=0$  and $W^{\sigma\sigma'}_{\rm B}(q \rightarrow \infty)\propto q^{-6}$, with the latter form being equivalent to the asymptotic behavior $S_{\uparrow\downarrow}(q\rightarrow \infty)\propto q^{-4}$. 

Finally, the Pauli principle requirement $g_{\sigma\sigma}(0)=0$ is ensured in our approach by the behavior of the Pauli potential in the origin, $v^{\sigma\sigma}_{\rm \scriptscriptstyle P}(r \rightarrow 0)\rightarrow 2~\hbar^2 /(m r^2)$. This behavior determines the power-law exponent in the distribution function of parallel-spin electrons, $g_{\sigma\sigma}(r \rightarrow 0) \propto r^2$, as can be proved directly from Eq. (5). 

\section{Numerical Results}
We turn to a presentation of our numerical results, which are obtained by solving Eq. (5) with the following self-consistency cycle. We start with the trial choice $g_{\sigma\sigma'}(r)=g^{\rm \scriptscriptstyle HF}_{\sigma\sigma'}(r)$ and $W^{\sigma\sigma'}_{\rm B}(r)=0$, and find the effective potentials $V_{\sigma\sigma'}(r)$ by means of Eq. (18) and hence the structure factors $S_{\sigma\sigma'}(q)$ via Eq. (20). At this point we can calculate new values for $g_{\sigma\sigma'}(r)$ and for $W^{\sigma\sigma'}_{\rm B}(r)$ by taking Fourier transforms and using Eq. (13). This procedure is repeated until self-consistency is achieved. The computational time typically needed to obtain $g_{\sigma\sigma'}(r)$ at each value of $r_s$ and $\zeta$ is a few minuted on a PC with a Pentium IV/1.4 GHz processor.

We have calculated in this way the spin-resolved pair distribution functions of a $3D$ electron gas for $r_s$ up to 100. The main results of our work are shown in Figures 1-8. 

In Figure 1 we show that our results for $g(r)=[g_{\uparrow\uparrow}(r)+g_{\uparrow\downarrow}(r)]/2$ in the paramagnetic EG at $r_s=1,5,10$ and $20$ are in excellent agreement with the QMC data of Ortiz {\it et al.}.\cite{ortiz} In the same range our results for $g(r)$ are in excellent agreement with those of Kallio and Piilo.\cite{kp} In Figures 2 and 3 we show that for the same cases our results for the spin-resolved pair functions $g_{\uparrow\uparrow}(r)$ and $g_{\uparrow\downarrow}(r)$ are also in excellent agreement with the QMC data of Ortiz {\it et al.}.\cite{ortiz} To the best of our knowledge theoretical results of similar quality have not been reported in the literature from an approach which is free of input and/or fitting parameters. Figure 4 shows a comparison of the present results at $r_s=10$ with those previously obtained by us\cite{gr} with a self-consistent Hartree approximation. 

In Figure 5 we compare our results for the paramagnetic EG at $r_s=40$ and $100$ with the QMC data of Ortiz {\it et al.}\cite{ortiz} and with those obtained more recently by Zong {\it et al.}.\cite{zong} As discussed by the latter authors, these two QMC approaches are essentially the same but involve some technical differences. In particular the study reported by Zong {\it et al.}\cite{zong} includes backflow and three-body terms in the wave function, and uses a novel numerical technique termed "twisted averaged boundary conditions" which allows a sizeable reduction of finite-size errors.\cite{private} The reader is referred to Ref. \onlinecite{zong} for details. These differences are not expected to change significantly the results of Ortiz {\it et al.}\cite{ortiz} for $r_s$ below $20$, but, as is clear from Figure 5, the differences between the two QMC studies become significant at larger $r_s$. Our results lie somewhere in between the two sets of QMC data at large $r_s$. From the theoretical point of view it is important to remark that in FHNC-type calculations at strong coupling the induced $W^{\sigma\sigma'}_{\rm B}(q)$ interactions should be corrected by the addition of three-body correlations and elementary-diagrams (or "bridge functions") contributions.\cite{lantto,zab,report} 

Further comparisons with QMC data on pair distribution functions can be made for the fully spin-polarized EG ($\zeta=1$). This is done in Figure 6 at $r_s=1,4,40$ and $100$. No QMC data are available in the intermediate range $4<r_s<40$, but the good agreement with the QMC data of Ortiz {\it et al.}\cite{ortiz} in the low-$r_s$ range is evident. The same inaccuracies in the theory that we have exposed in the paramagnetic EG at strong coupling are also found at $\zeta=1$. 

Finally, in Figure 7 we report the spin-dependent effective electron-electron interaction $V_{\rm eff}^{\sigma\sigma'}(r)=v(r)+W^{\sigma\sigma'}_{\rm B}(r)+\delta_{\sigma\sigma'}W^{\sigma\sigma}_{\rm e}(r)$ and its Fourier transform $V_{\rm eff}^{\sigma\sigma'}(q)$, 
as it emerges from our self-consistent calculations on a $3D$ EG both in the paramagnetic and in the ferromagnetic state. The attractive part of the parallel-spin effective interaction deepens at increased $\zeta$ as is physically expected. The Pauli potential and its dependence on $\zeta$ are shown in Figure 8. 

\subsection{Ground-state energy}
The ground-state energy $\varepsilon_{\rm g}$ (per electron) of the EG at each value of $r_s$ and $\zeta$ 
can be calculated by means of an integration over the coupling constant $\lambda$,
\begin{equation}\label{e12}
\varepsilon_{\rm g}=\varepsilon_0+\frac{1}{2}\int_{0}^{1} \frac{d\lambda}{\lambda}~\int \frac{d^3 {\bf q}}{(2 \pi)^3}~v^{(\lambda)}_q~[S_{\lambda}(q)-1]
\end{equation} 
directly from the structure factor $S_{\lambda}(q)$ calculated with interaction $v^{(\lambda)}_q=4 \pi e^2\lambda/q^2$. 
Here $\varepsilon_0=3[(1+\zeta)^{5/3}+(1-\zeta)^{5/3}]/(10 \alpha^2 r^2_s)~{\rm Ryd}$ 
is the ideal-gas kinetic energy with $\alpha=(9 \pi/4)^{-1/3}$, and $S(q)$ has been defined immediately below Eq. (24). In fact, the integration over $\lambda$ is carried out by integration over $r_s$. The second term in Eq. (25) is the exchange-correlation (xc) energy, from which the correlation energy is obtained by subtracting the expression for the exchange energy  $\varepsilon_{\rm x}=-3[(1+\zeta)^{4/3}+(1-\zeta)^{4/3}]/(4\pi \alpha r_s)\,{\rm Ryd}$. 	

We have calculated the ground-state energy for $\zeta=0,0.333,0.667$ and $1$ over the range $1 \leq r_s \leq 50$. The results are reported in Table I in comparison with QMC data from Ceperley and Alder,\cite{ceperley} Ortiz {\it et al.}\cite{ortiz} and Zong {\it et al.}.\cite{zong} 
The Table also includes other theoretical results obtained in the self-consistent dielectric theory of Singwi {\it et al.}\cite{stls} and in the modified-convolution approximation of Tanaka and Ichimaru.\cite{tanaka} 	

It is seen from Table I that the present theoretical approach yields fairly accurate values of the ground-state energy even at $r_s\simeq 50$, even though the details of the pair distribution function are becoming quantitatively inaccurate as we have already seen in Figures 5 and 6. Much higher accuracy is needed for theoretical predictions on the quantum phase transition from the paramagnetic phase to the fully spin-polarized phase, since from the QMC study of Zong {\it et al.}\cite{zong} the difference in energy between these two phases at $r_s=100$ is only about $4\times 10^{-6} {\rm Ryd}$ in favor of the latter. A continuous transition between these two states seems to start for $r_s \simeq 50$-$70$ according to the QMC data. In contrast, the theoretical approaches of Singwi {\it et al.}\cite{stls} and of Tanaka and Ichimaru\cite{tanaka} suggest that the magnetically ordered state may be the favored one already at $r_s\simeq 40$-$50$. 
\section{Summary and Discussion}

In summary, we have presented in this work a new theoretical study of the spin-resolved pair distribution functions and of the ground-state energy of the $3D$ electron gas. Our approach yields numerical results of good quality in the regime of weak and intermediate coupling strength. 	

Improvements of the theory will be necessary for a quantitative study of the magnetic phase diagram and for an extension to electron fluids of lower dimensionality. As we have already commented in the main text, the kinetic energy functional may be treated more accurately by recourse to an analysis of the pair functions into Kohn-Sham pair orbitals.\cite{gr,dbose} Preliminary calculations indicate that an appreciable improvement can be obtained in this way in the values of the ground-state energy at small to intermediate coupling over the values reported in Table I. At larger values of the coupling we expect that a sizeable improvement on the FHNC approximation to the effective electron-electron interactions can only arise from the inclusion of contributions from the so-called elementary diagrams and from three-body correlations.\cite{report} Better quantitative accuracy can be expected to be achieved by such means when the self-consistency requirements on the theory are extended to include the thermodynamic sum rules on the compressibility and spin susceptibility of the electron fluid.\cite{dpt}

\begin{acknowledgements}
This work was partially supported by MIUR through the PRIN2001 Initiative. 
We are indebted to Prof. P. Ballone and Prof. D. M. Ceperley 
for providing us with their QMC data reported in the Figures.
\end{acknowledgements}

\newpage

\begin{table}
\caption{Energy of the $3D$ EG in Ryd/electron. 
CA/QMC from Ceperley and Alder,~\cite{ceperley} OHB/QMC from Ortiz {\it et al.},~\cite{ortiz}
ZLC/QMC from Zong {\it et al.}.~\cite{zong}
STLS from Singwi {\it et al.}.~\cite{stls} and TI from Tanaka and Ichimaru.~\cite{tanaka}}
\begin{tabular}{lllllllll}
\hline\hline
$r_s$ &Various calculations&$\zeta=0.0$&$\zeta=0.333$&$\zeta=0.667$&$\zeta=1.0$\\ \hline
1			&CA/QMC  							 &1.174&&&\\
			&OHB/QMC							 &1.181&&&2.294\\
			&Present work	 			 	 &1.17810&1.29786&1.66346&2.29753\\
			&STLS 							   &1.1704&1.2885&1.6505&2.2849\\
			&TI 							   	 &1.167&&&2.281\\ \hline
5			&CA/QMC  							 &-0.1512&&&-0.1214\\
			&Present work	 			 	 &-0.14974&-0.14616&-0.13557&-0.11846\\
			&STLS 							   &-0.1511&-0.1483&-0.1393&-0.1235\\
			&TI 							   	 &-0.1544&&&-0.1267\\ \hline 
10		&CA/QMC  							 &-0.107&&&-0.101\\
			&OHB/QMC							 &-0.106&&&-0.101\\
			&Present work	 			 	 &-0.10562&-0.10488&-0.10277&-0.09957\\
			&STLS 							   &-0.1058&-0.1055&-0.1043&-0.1020\\ 
			&TI 							   	 &-0.1084&&&-0.1046\\ \hline  
20		&CA/QMC  							 &-0.063&&&-0.0625\\
			&OHB/QMC							 &-0.063&&&-0.0625\\
			&Present work	 			 	 &-0.06265&-0.06250&-0.06210&-0.06153\\
			&STLS 							   &-0.0623&&&-0.0623\\ 
			&TI 							   	 &-0.0642&&&-0.0642\\ \hline  
40		&ZLC/QMC &-0.03523748(60)&-0.03523295(67)&-0.03520539(67)&-0.03513483(72)\\
			&Present work	 			   &-0.03470&-0.03467&-0.03459&-0.03450\\
			&STLS 							   &-0.0342&&&-0.0345\\ \hline	 
50		&ZLC/QMC&-0.02889900(62)&-0.02889962(68)&-0.02888835(62)&-0.02884983(81)\\
			&OHB/QMC							 &-0.029&&&-0.0288\\
			&Present work	         &-0.02845&-0.02844&-0.02839&-0.02834\\
			&STLS                  &-0.0280&&&-0.0282\\ 
			&TI 							   	 &-0.0290&&&-0.0293\\ \hline \hline
\end{tabular}
\end{table}

\newpage
\begin{figure}
\label{f1}
\begin{center}
\includegraphics[scale=0.6]{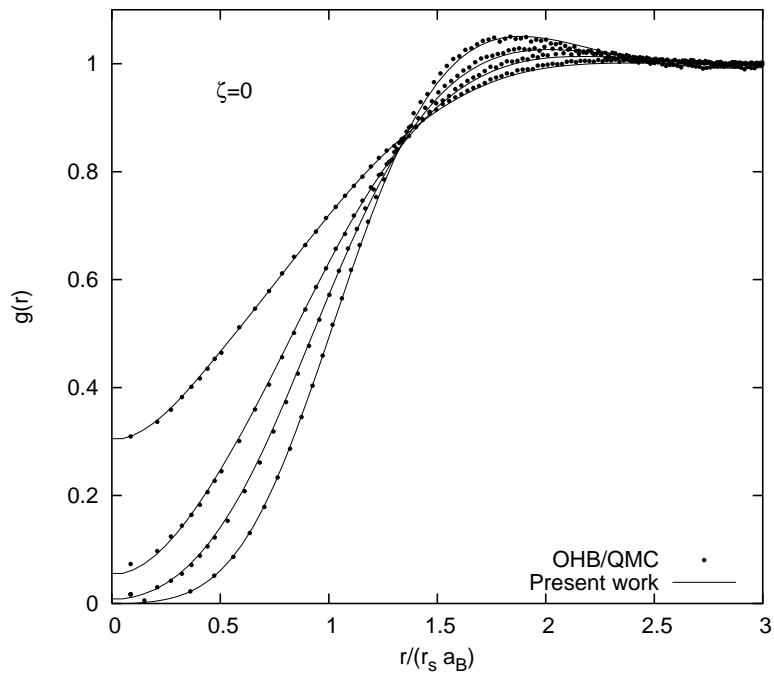}
\caption{The pair distribution function $g(r)$ in a paramagnetic $3D$ EG at $r_s=1,5,10$ and $20$ (from top to bottom at low $r$), as a function of $r/(r_s a_B)$. The results of the present work (full lines) are compared with QMC data of Ortiz {\it et al.}~\cite{ortiz} (dots).}
\end{center}
\end{figure}

\begin{figure}
\label{f2}
\begin{center}
\includegraphics[scale=0.6]{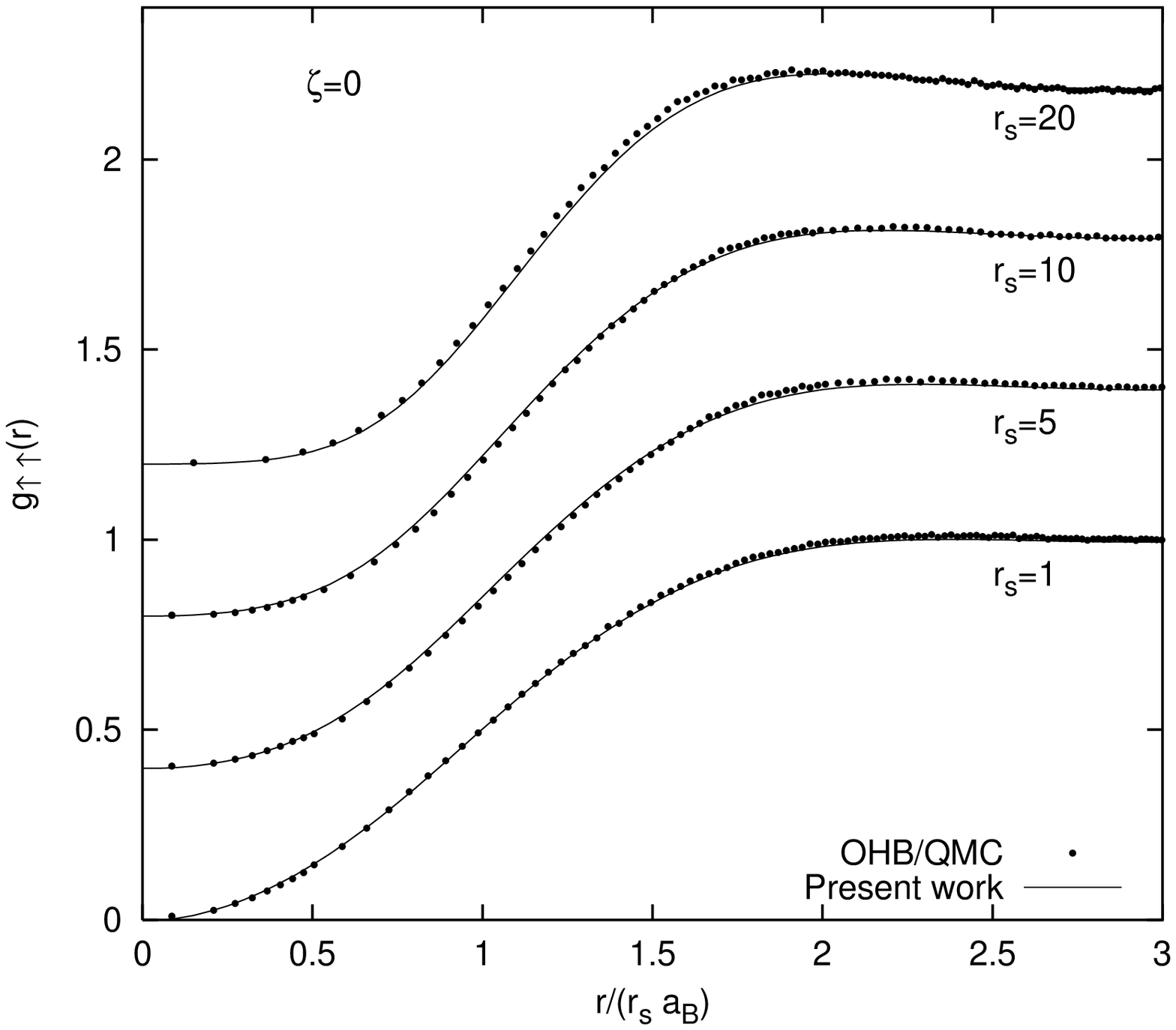}
\caption{The parallel-spin pair distribution function $g_{\uparrow\uparrow}(r)$ in a paramagnetic $3D$ EG at $r_s=1,5,10$ and $20$, as a function of $r/(r_s a_B)$. The results of the present work (full lines) are compared with QMC data of Ortiz {\it et al.}~\cite{ortiz} (dots). The curves at $r_s=5,10$ and $20$ have been shifted upwards for clarity by 0.4, 0.8 and 1.2, respectively.}
\end{center}
\end{figure}

\newpage

\begin{figure}
\begin{center}
\includegraphics[scale=0.6]{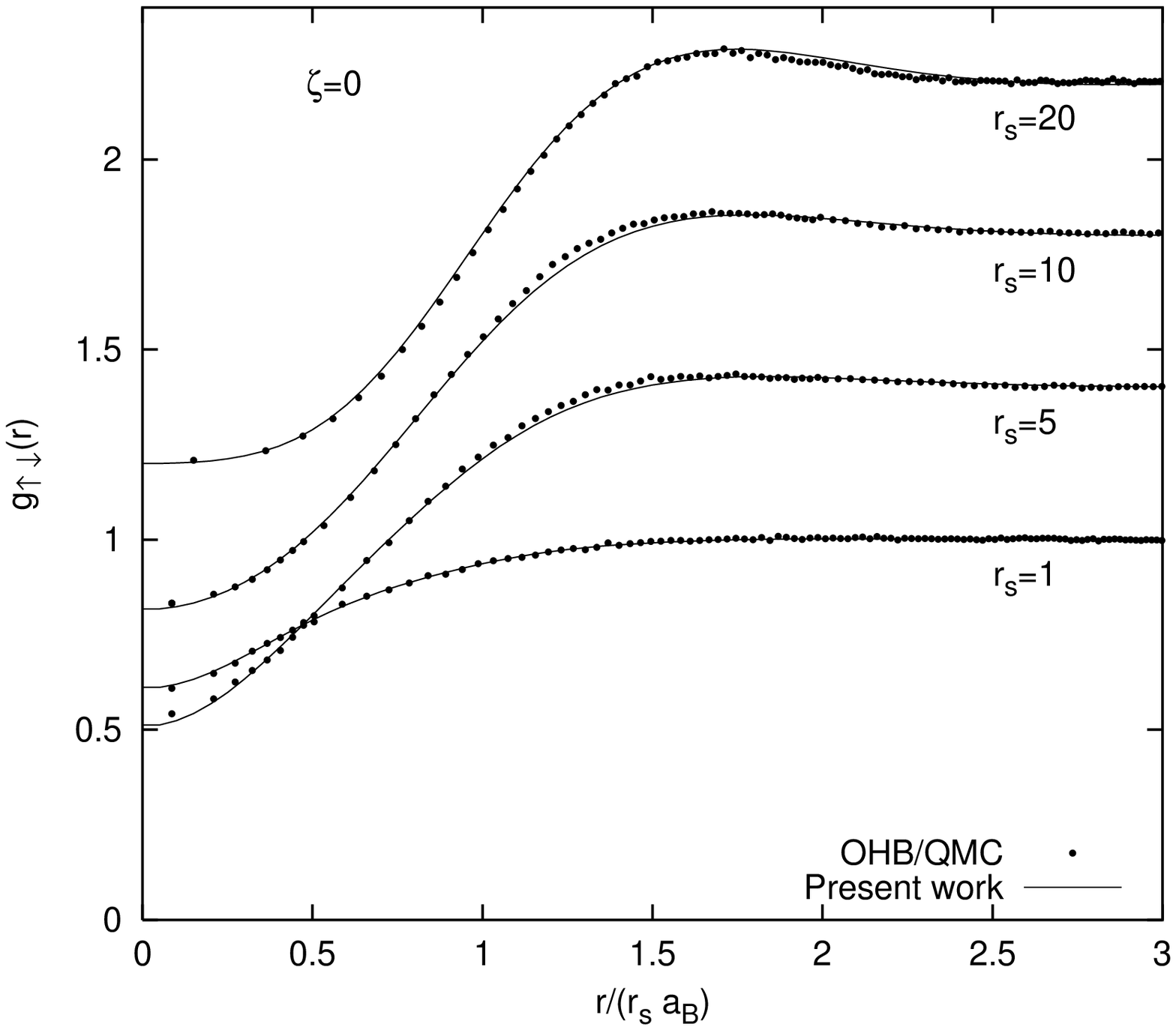}
\caption{The antiparallel-spin pair distribution function $g_{\uparrow\downarrow}(r)$ in a paramagnetic $3D$ EG at $r_s=1,5,10$ and $20$, as a function of $r/(r_s a_B)$. The results of the present work (full lines) are compared with QMC data of Ortiz {\it et al.}~\cite{ortiz} (dots). The curves at $r_s=5,10$ and $20$ have been shifted upwards by 0.4, 0.8 and 1.2, respectively.}
\end{center}
\label{f3}
\end{figure}

\begin{figure}
\begin{center}
\includegraphics[scale=0.6]{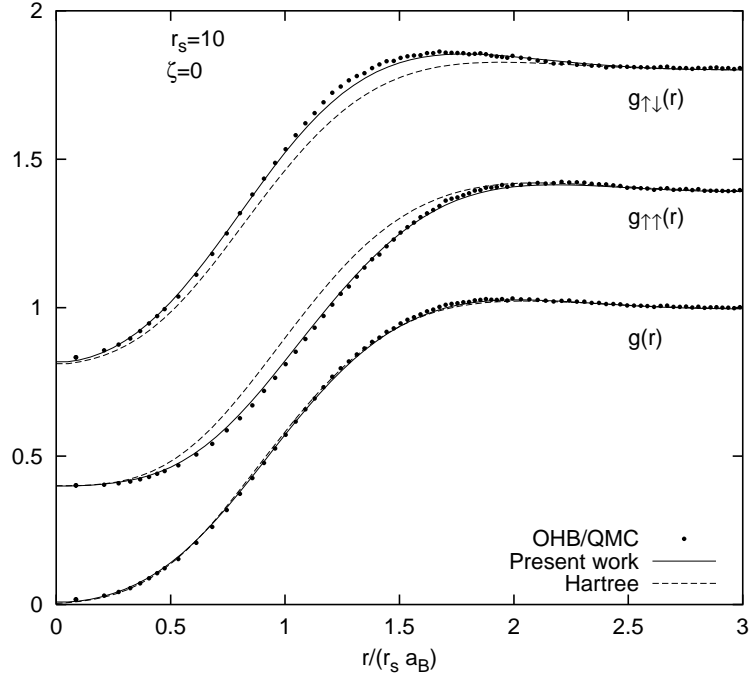}
\caption{The average $g(r)$ and spin-resolved pair functions in a paramagnetic $3D$ EG at $r_s=10$, as a functions of $r/(r_s a_B)$. The results of the present work (full lines) are compared with our previous Hartree results~\cite{gr} (dashes) and with QMC data of Ortiz {\it et al.}~\cite{ortiz} (dots). The results for $g_{\uparrow\uparrow}(r)$ have been shifted upwards by 0.4 and that for $g_{\uparrow\downarrow}(r)$ by 0.8.}
\end{center}
\label{f4}
\end{figure}

\newpage

\begin{figure}
\begin{center}
\includegraphics[scale=0.6]{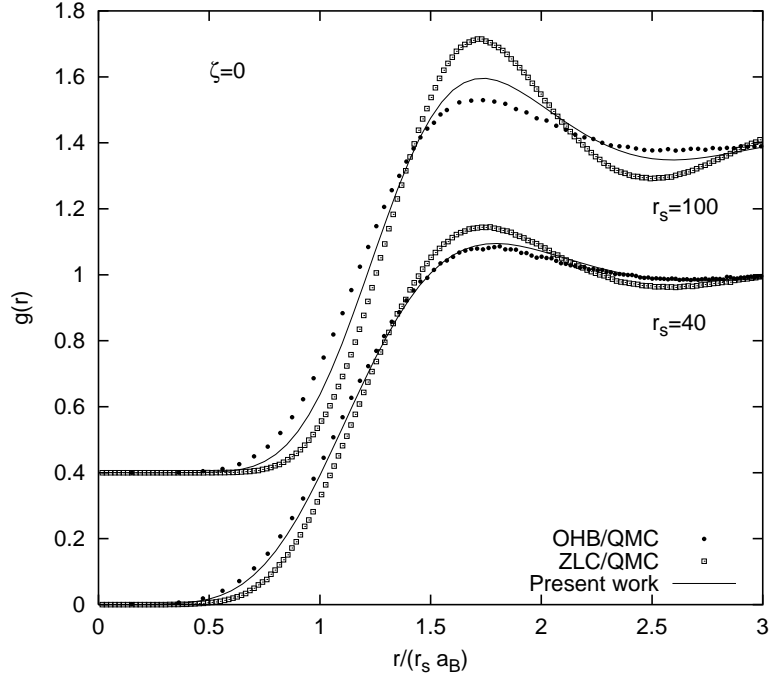}
\caption{The pair distribution function $g(r)$ in a paramagnetic $3D$ EG at $r_s=40$ and $100$, 
as a function of $r/(r_s a_B)$. The results of the present work (full lines) are compared with QMC data of Ortiz {\it et al.}~\cite{ortiz} (dots) and of Zong {\it et al.}~\cite{zong} (empty boxes). The curves at $r_s=100$ have been shifted upwards by 0.4.}
\end{center}
\label{f5}
\end{figure}

\begin{figure}
\begin{center}
\includegraphics[scale=0.6]{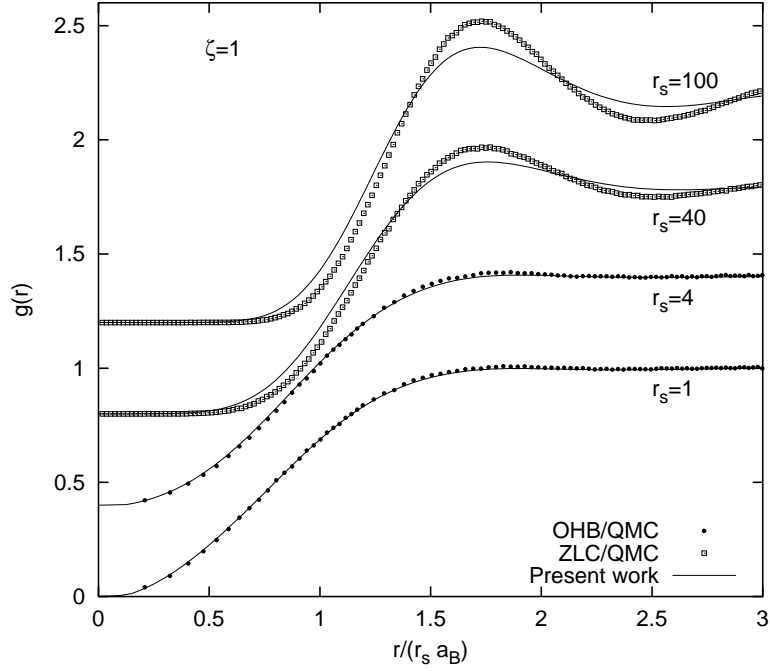}
\caption{The pair distribution function $g(r)$ in the fully spin-polarized $3D$ EG at $r_s=1,4,40$ and $100$, as a function of $r/(r_s a_B)$. The results of the present work (full lines) are compared with QMC data of Ortiz {\it et al.}~\cite{ortiz} (dots) and of Zong {\it et al.}~\cite{zong} (empty boxes). The curves at $r_s=4,40$ and $100$ have been shifted upwards by 0.4, 0.8 and 1.2, respectively.}
\end{center}
\label{f6}
\end{figure}
\newpage

\begin{figure}
\begin{center}
\includegraphics[scale=0.5]{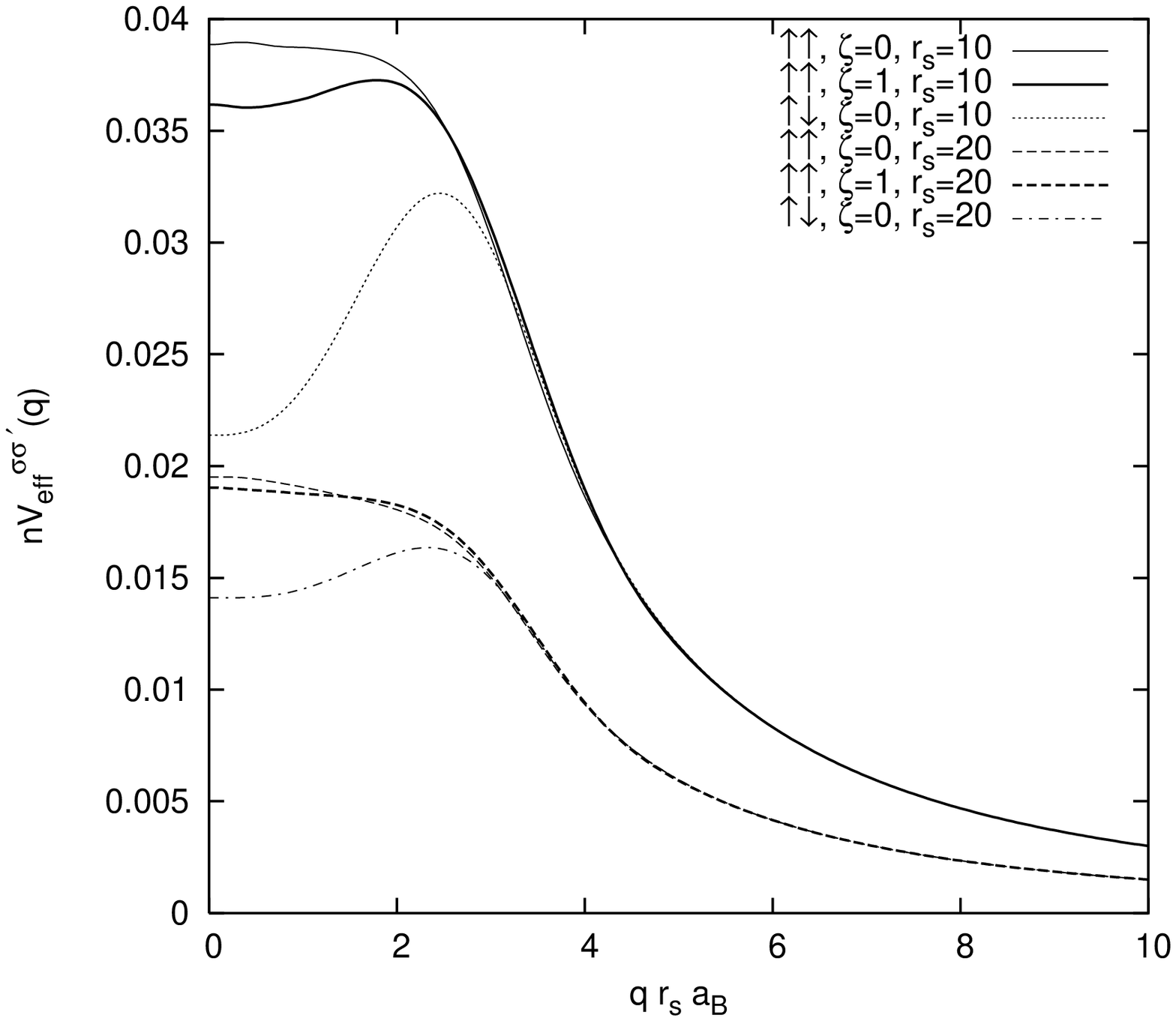}
\includegraphics[scale=0.5]{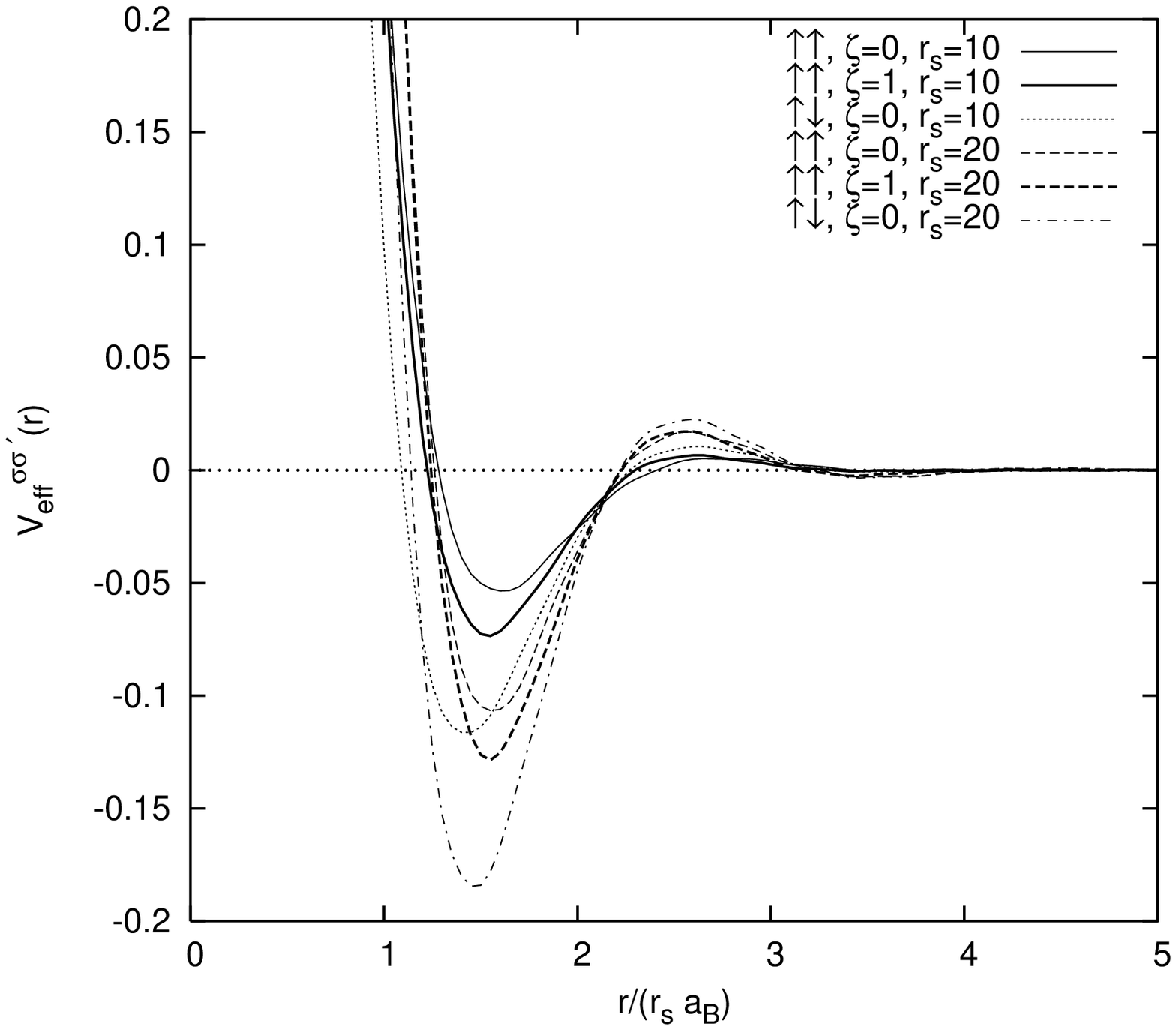}
\caption{Left Panel (momentum space): the spin-dependent effective interaction $n V_{\rm eff}^{\sigma\sigma'}(q)$ (in units of $e^2/a_B$) in a $3D$ EG at $r_s=10$ and $20$, as a function of $q r_s a_B$. Right Panel (real space): the spin-dependent effective interaction $V_{\rm eff}^{\sigma\sigma'}(r)$ (in units of $\hbar^2 k_F^2/m$) in a $3D$ EG at $r_s=10$ and $20$, as a function of $r/(r_s a_B)$.}
\end{center}
\label{f7}
\end{figure}

\begin{figure}
\begin{center}
\includegraphics[scale=0.6]{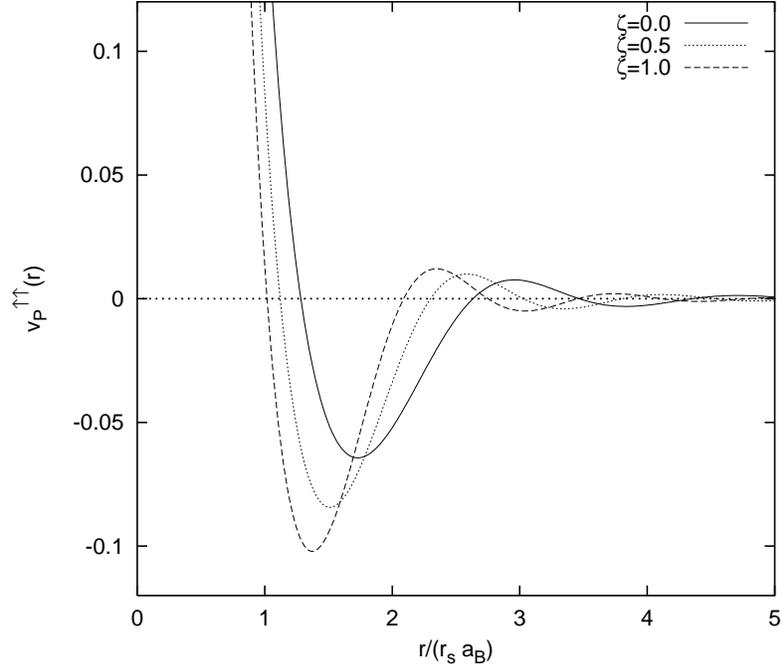}
\caption{The Pauli potential $v^{\sigma\sigma'}_{\rm \scriptscriptstyle P}(r)$ (in units of $\hbar^2 k_F^2/m$), as a function of $r/(r_s a_B)$. With this choice of units the Pauli potential has no explicit dependence on $r_s$.}
\end{center}
\label{f8}
\end{figure}
\end{document}